# Serum protein resistant behavior of multisite-bound poly(ethylene glycol) chains onto iron oxide surfaces

Nicoletta Giamblanco[1*], Giovanni Marletta[1*], Alain Graillot[2], Nicolas Bia[2], Cédric Loubat[2] and Jean-François Berret[3]*

[1]Laboratory for Molecular Surface and Nanotechnology (LAMSUN), Department of Chemical Sciences, University of Catania and CSGI, Viale A. Doria 6, 95125, Catania, Italy

[2] Specific Polymers, ZAC Via Domitia, 150 Avenue des Cocardières, 34160 Castries, France

[3]Matière et Systèmes Complexes, UMR 7057 CNRS Université Denis Diderot Paris-VII, Bâtiment Condorcet, 10 rue Alice Domon et Léonie Duquet, 75205 Paris, France.

**Abstract:** Recent surveys have shown that the number of nanoparticle-based formulations actually used at the clinical level is significantly lower than expected a decade ago. One reason for this is that the nanoparticle physicochemical properties fall short for handling the complexity of biological environments and for preventing nonspecific protein adsorption. In this study, we address the issue of the interactions of plasma proteins with polymer coated surfaces. To this aim, we use a non-covalent grafting-to method to functionalize iron oxide sub-10 nm nanoparticles and iron oxide flat substrates, and compare their protein responses. The functionalized copolymers consist in alternating poly(ethylene glycol) (PEG) chains and phosphonic acid grafted on the same backbone. Quartz Crystal Microbalance with dissipation was used to monitor the polymer adsorption kinetics and to evaluate the resistance to protein adsorption. On flat substrates, functionalized PEG copolymers adsorb and form a brush in the moderate or in the highly stretched regimes, with density between 0.15 and 1.5 nm$^{-2}$. PEG layers using phosphonic acid as linkers exhibit excellent protein resistance. In contrast, layers prepared with carboxylic acid as grafting agent exhibit mitigated protein responses and layer destructuration. The present study establishes a correlation between the long-term stability of PEG coated particles in biofluids and the protein resistance of surfaces coated with the same polymers.

Corresponding author: jean-francois.berret@univ-paris-diderot.fr

## I - Introduction

In nanomedicine, the possibility to use engineered nanoparticles for medical imaging and therapy has attracted much interest during the last 15 years. Recent surveys have shown however that nanotechnology-based formulations have not been as successful as initially thought.[1] The number of nanoparticle carriers actually used to improve patient outcomes at the clinical level is significantly lower than expected a decade ago. Today most therapeutic drug-carrying particles are in the form of liposomes, lipid-based complexes, or biodegradable polymer/drug combinations. More complex nano-formulations, e.g. including inorganic particles have been barely exploited or are still in clinical trials. One of the reasons of these mixed results is related to the difficulty to match the bare physicochemical properties of nanoparticle carriers to the constraints of the biological environments, and in particular to prevent the ubiquitous nonspecific protein adsorption. For a vast majority, particles administered *in vivo* are recognized by plasma proteins and eliminated from the blood stream within a few minutes, leading to their accumulation in unrelated organs such as liver and kidneys.[2,3]

On the physicochemical side, it is now well established that for biological applications, nanoparticle surfaces need to be modified to prevent protein adsorption. A great deal of surface functionalization methods has been developed and assessed either *in vitro* or *in vivo*. The most advanced strategies have been based on modified or grafted polymers,[4-11] although new fabrication techniques using supported bilayer or biomembrane mimetics are currently evaluated.[12] With polymers, surface functionalization is achieved either by physical adsorption methods (such as spin-coating, layer-by-layer assembly, solvent-casting).[13,14] – or by chemical bonding methods based on surface-initiated polymerization[10,11] or on surface activation by means of radiation treatments.[15-17] Surface functionalization takes advantage of the extended library of polymer architectures (linear chains, copolymers, stars, dendrimers) and of chelating agents that were developed in polymer and coordination chemistry.[18-24] Hydrosoluble neutral polymers such as poly(ethylene glycol) (PEG), poly(acrylamide) and some polysaccharides show an improved protein resistance (as compared to ion-containing chains) and were incorporated into nanomedicine formulation synthesis. Among the polymers tested, PEG is the most studied bioresistant polymer.[3,21,23-33] Apart from being inexpensive and approved by regulatory and control agencies, poly(ethylene glycol) offers many advantages. Made of a sequence of –CH$_2$–CH$_2$–O– monomers, PEG is a flexible macromolecule and can be synthesized with narrow molecular weight dispersity. Moreover, PEG was also found to follow accurately polymer dynamics predictions[34,35] and by so doing it allows quantitative evaluation of the chain conformation, in view of the importance of this factor in determining the adsorption protein behavior.

For polymers at curved or flat interfaces, chain conformations may be strikingly different. Alexander and de Gennes were the first to describe theoretically the conformational behavior of polymers at interfaces, and specifically polymers tethered by one extremity and having the remaining part of the chain dangling in the solvent.[36,37] For flat surfaces, two main regimes were predicted.[37,38] At low polymer density $\sigma$ such $\sigma < 1/\pi R_g^2$, where $R_g$ is the gyration radius of the chain in good solvent conditions, the polymers adopt a so-called mushroom configuration. In this first case, the adlayer thickness is twice the gyration radius. At higher densities, $\sigma > 1/\pi R_g^2$, monomer-monomer excluded volume interactions induce a stretching of the chains, which then enter into the brush regime. In this configuration, the height increases and varies as $h \sim \sigma^\nu N$, where $N$ is with the degree of polymerization and $\nu$ a coefficient between 1/3 and 1 that depends on the solvent quality.[38-40] It is commonly admitted that the soft interfaces represented by hydrosoluble polymer brushes are excellent protein repellent.[6,15,21,25,27,30,41-43] Polymer adlayers or brushes are generally regarded as steric repulsive barriers. Recent studies have shown however that protein interactions with soft interfaces are far more complex and that the brushes act as kinetic barriers rather than efficient prevention of adsorption.[42]

The present report aims to establish a correlation between the stability of PEGylated particles in biological environments, and the protein resistance of PEGylated surfaces coated with the same polymers. To this end, we used a non-covalent grafting-to method to deposit functionalized PEG copolymers on iron oxide substrates under environmentally friendly conditions, *i.e.* in aqueous media and at room temperature. Copolymers studied consist in alternating PEG chains (of molecular weight 1000, 2000 and 5000 g mol$^{-1}$) and acidic moieties grafted on the same backbone. The deposition on iron oxide is driven by the acid groups, which are of two kinds, carboxylic acid and phosphonic acid. Phosphonic acid is known to have a higher affinity towards metals or metal oxides compared to sulfates and carboxylates, and it is anticipated that these residues will build stronger links with the surface.[21-23,32,41,44,45] This research follows up recent pharmacokinetics studies showing that sub-10 nm iron oxide nanoparticles coated with the above polymers circulate in the blood pool for over than 2 hours without being recognized by the mononuclear phagocytic system.[24] The circulation time was about 50 times larger than that of non-PEGylated probes and benchmarks, a feature that was unequivocally attributed to the coating. To study the interaction of PEGylated polymers and proteins with iron oxide substrates, Quartz Crystal Microbalance with Dissipation (QCM-D) was carried out with a twofold objective: *i)* monitor the polymer adsorption kinetics and derive the adlayer structure, and *ii)* evaluate the protein resistance of the built-up layers. By varying parameters such as the acid-base conditions, the polymer molecular weight and the nature of the grafting agent, different protein resistance performances were obtained, with values ranging typically from 65% to 99 % of repulsion efficacy. Interestingly, we have found a positive correlation between the strength of the PEG-substrate linkage, the stability of PEGylated nanoparticles and the protein resistance determined by QCM-D.

## II – Experimental Section
**Polymer synthesis and characterization**

*Synthesis:* Poly(ethylene glycol methacrylate-*co*-dimethyl(methacryoyloxy) methyl phosphonic acid), abbreviated phosphonic acid PEG copolymer in the following was synthesized by Specific Polymers®, France (http://www.specificpolymers.fr/). The synthesis was carried out by free radical polymerization from PEG-methacrylate (PEGMA, SP-43-3-002, CAS: 26915-72-0) and dimethyl(methacryoyloxy)methyl phosphonate (MAPC1, SP-41-003, CAS: 86242-61-7) monomers, leading to the poly(PEGMA-*co*-MAPC1) statistical polymer (Supporting Information S1). PEGMA and MAPC1 monomer conversion rates were determined during synthesis and showed similar time dependences, indicating that the copolymers have the same number of PEGs and of phosphonic acid groups (Supporting Information S2). For this synthesis, the molecular weights of the PEG pending side-chains were 1000, 2000 and 5000 g mol$^{-1}$, refereed to as PEG$_{1K}$, PEG$_{2K}$ and PEG$_{5K}$ in the sequel of the paper (Fig. 1a). Poly(poly(ethylene glycol) methacrylate-co-methacrylic acid), in short poly(PEGMA-co-MAA) was synthesized by free radical polymerization from PEG-methacrylate and from methacrylic acid (MAA) (CAS: 79-41-4, Acros Organics) monomers (Supporting Information S2). The resulting statistical copolymer is abbreviated carboxylic acid PEG copolymer. For this synthesis the molecular weight of PEG pending side chains studied is 2000 g mol$^{-1}$.

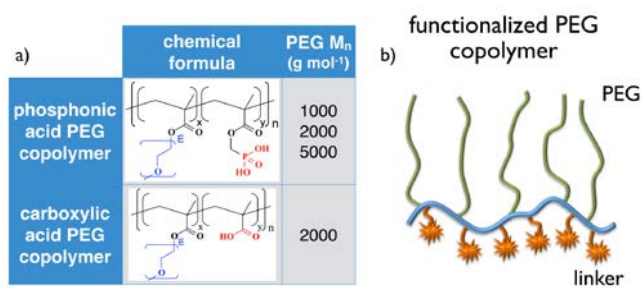

**Figure 1**: a) Chemical formulae of the statistical copolymers used in this work. Four copolymers were investigated, three with phosphonic acid and PEG side chains 1000, 2000 and 5000 g mol$^{-1}$, and one with carboxylic acid and PEG 2000 g mol$^{-1}$. b) Schematic representation of the statistical copolymers

*Characterization by light scattering:* The polymer weight-averaged molecular weight was determined by static light scattering measurements (NanoZS Zetasizer, Malvern). Toluene was used for calibration. The polymers solutions were prepared with 18.2 MΩ MilliQ water, filtered with 0.2 μm cellulose filters and their $pH$ was adjusted to 8 by addition of sodium hydroxide. The scattering intensity was found to vary linearly with the concentration between 0 and 1 wt. % (Supporting Information S2). The refractive index increment of the different polymer solutions was obtained from refractometry (Arago, Cordouan Technologies), and it was used to calculate the polymer scattering contrast. The molecular weights were derived from the Zimm representation and from the zero-concentration extrapolated scattering, as detailed in Ref.[21,32]. The weight-averaged molecular weights of the 4 polymers studied are provided in Table 1. They are in good agreement with the targeted ones.

For the characterization of the iron oxide particles, dynamic light scattering was performed using the NanoZS at the wavelength of 633 nm and in a close-to-backscattering configuration, *i.e.* with a scattering angle of 173°. From the time dependence of the scattered intensity, the second-order autocorrelation function of the light was calculated and analyzed using the cumulant method and the CONTIN algorithm. Both procedures gave consistent values for the hydrodynamic diameter $D_H$. For bare nanoparticles the hydrodynamic diameters were found to be larger than those obtained by TEM, the reason being attributed to the particle size distribution and to the fact that the scattered intensity varies as the sixth power of the particle diameter.

| terminus | $M_n$(PEG) g mol$^{-1}$ | $M_n$ g mol$^{-1}$ | $M_w$ g mol$^{-1}$ | acid group meq g$^{-1}$ | funct. group per chain |
|---|---|---|---|---|---|
| phosphonic acid | 1000 | 5300 | 9500 | 1.58 | 4.2 |
|  | 2000 | 7200 | 12950 | 0.87 | 3.1 |
|  | 5000 | 42600 | 76650 | 0.35 | 7.4 |
| carboxylic acid | 2000 | 8600 | 15500 | 0.46 | 4.0 |

**Table 1**: Structural parameters of the phosphonic and carboxylic PEG copolymers used in this work. $M_n$(PEG) denotes the PEG molecular weight of the pendant side chains, $M_n$ and $M_w$ the number and mass averaged molecular weights of the statistical copolymers as determined by light scattering. The molar equivalent of acid groups per gram (milli eq g$^{-1}$) of polymer was determined from $^1$H and $^{31}$P NMR, leading to the number of functionalized groups per chain.

*Acid-base titration:* To study the role of the acid functionality on grafting, poly(PEGMA-co-MAPC1) and poly(PEGMA-co-MAA) adsorptions were performed at two pH values, pH 2.0 and pH 7.4. pH 2.0 corresponds to the conditions for coating iron oxide nanoparticles.[21,24,46] Acid-base titration curves have shown the presence of two $pK_A$'s ($pK_{A1}$ = 2.7 and $pK_{A2}$ = 7.8) for the phosphonic acid (Supporting Information S3). For carboxylic acid, the $pK_A$ was found at 5.5. For iron oxides, the point of zero charge (PZC) is observed around pH 8.0. Below, the surface is positively charged due to Fe-OH$_2^+$ groups whereas above the surface bear Fe-O$^-$ negative charges.[47,48] It should be mentioned here that the iron oxide nanoparticles tested are made from maghemite γ-Fe$_2$O$_3$, whereas the substrate is magnetite, Fe$_3$O$_4$. As shown by Jolsterå et al. using high precision potentiometric titrations,[49] the acid/base properties of magnetite are similar to those of maghemite, except for the difference in surface site density, estimated at 1.50 nm$^{-2}$ for magnetite, and 0.99 nm$^{-2}$ for maghemite. For this reason, we assume that the two iron oxide substrates behave similarly with respect to the acidic residues.

*Characterization by size exclusion chromatography:* The molar-mass dispersity for poly(PEGMA-co-MAPC1) and poly(PEGMA-co-MAA) was determined from size exclusion chromatography on PolyPore column using THF as eluent and polystyrene standards. For PEG$_{2K}$ side chains, it was found at 1.81 and 1.78 respectively.[21]

*Characterization by NMR:* Phosphonic acid PEG copolymers were characterized by $^1$H NMR and $^{31}$P NMR using a Bruker Avance 300 spectrometer operating at 300 MHz (Supporting Information S4). From the molar equivalent of acid groups per gram obtained by NMR, the average number of functional moieties and PEG side chains per chain was derived. It was found to be between 3 and 7 depending on the PEG side chain molecular weight (Table 1). These findings confirm the existence of multiple functional groups on the same polymer backbone (Fig. 1b).

*Iron oxide nanoparticles:* Iron oxide nanocrystals with diameter 6.8 nm were synthesized by co-precipitation of iron(II) and iron(III) salts in aqueous media and by further oxidation of the magnetite (Fe$_3$O$_4$) into maghemite (γ-Fe$_2$O$_3$).[46,50,51] As-prepared particles are positively charged and have nitrate counterions at their surfaces. The resulting interparticle interactions are repulsive, and provide long-term excellent colloidal stability to the dispersion. The particle size distribution was determined from dynamic light scattering and transmission electron microscopy. Their crystalline cubic structure was assessed by electron beam microdiffraction (Supporting Information S5).[52]

*Nanoparticle coating:* For nanoparticles coating, we used a protocol established in 2008[32,46] and later applied for the development of MRI contrast agents for *in vivo* experimentations.[24] Dispersions of particles and PEGylated copolymers were prepared in the same conditions of pH (pH 2.0) and concentration, and then mixed at different volume ratios $X$. The choice of pH was dictated by the fact the uncoated iron oxide particles are aggregating in neutral or alkaline conditions. Following the mixing, the polymers adsorb spontaneously at the particle surfaces due to the acid groups complexing the iron hydroxide sites on the magnetite surface,[32] resulting in an increase of the particle hydrodynamic diameter (Fig. 2a). With the PEGylated copolymers studied here, no precipitation or particle aggregation was observed following the mixing at this pH. The pH of the mixed solution was then raised to pH 8.0 by sodium hydroxide addition, leading to two distinct behaviors. Below a critical mixing ratio $X_C$, well-dispersed coated particles were obtained, again with a $D_H$ slightly larger than that of bare particles. Above $X_C$, particles precipitate and form large clumps in solution, which eventually sediment. Detailed investigations on polymer coverage as a function of the mixing ratio are discussed in Refs.[21,24]

*Serum proteins:* Phosphate buffer solution (PBS) was prepared by dissolving one tablet (Sigma-Aldrich) in 200 mL of de-ionized water (Millipore 18.2 MΩ resistivity), resulting in a solution ionic strength of 10 mM for the phosphate salts, 2.7 mM for potassium chloride, and 137 mM for sodium chloride (pH 7.4 at 25 °C). Fetal bovine serum (FBS) from Gibco invitrogen, with nominal composition 23.9 g L$^{-1}$ of BSA, α-globulin 13.2 g mL-1, β-globulin 4.5 g L$^{-1}$ and γ-globulin 0.155 g L$^{-1}$ was used. The FBS was diluted with PBS at concentration of 10 vol. %, and used without further modification.

*Nanoparticle stability in physiological and culture media:* For the evaluation of the particle stability, the following protocol was applied.[46,53] A few microliters of a concentrated dispersion were poured and homogenized rapidly in 1 mL of the solvent to be studied, and simultaneously the scattered intensity $I_S$ and diameter $D_H$ were measured by light scattering. After mixing, the measurements were monitored over a 2-hour period, and subsequent measurements were made after 24 hours and 1 week. Nanoparticles are considered to be stable if their hydrodynamic diameter $D_H$ in a given solvent remains constant as a function of the time and equal to its initial value. Solvents surveyed here are DI-water (pH 7.4), phosphate buffer saline and cell culture medium (Dulbecco's Modified Eagle's Medium, DMEM) with or without fetal bovine serum. Stability assays were performed on bare and coated 6.8 nm iron oxide nanoparticles. Examples of temporal behaviors in physiological and cell culture media together with the particle stability diagram are shown in Fig. S6.

**Quartz Crystal Microbalance**

A Quartz Crystal Microbalance with dissipation monitoring equipment (Q-Sense E1 system, Biolin Scientific, Sweden) was used to follow the adsorption kinetics of phosphonic acid and carboxylic acid PEG copolymers on iron oxide substrate. AT-cut quartz crystal sensors coated by a thin film of magnetite ($Fe_3O_4$) (Biolin Scientific, Sweden), with a fundamental resonance frequency of 4.95 MHz, were cleaned by 10 min sonication and exposed to an ultraviolet (UV)/ozone cleaner for 10 minutes. QCM-D experiments were carried out at 25 ± 0.02 °C in an exchange mode at a flow rate of 100 µL min$^{-1}$. Injected polymer solutions were prepared at concentration 0.1 wt. % and at pH 2.0 and pH 7.4 adjusted with addition of sodium hydroxide or hydrochloric acid. At least 0.5 mL of the sample solution was delivered into the chamber containing the crystal sensor (of internal volume 40 µL) to ensure a complete liquid exchange. In a typical experiment the crystal was excited at its fundamental resonance frequency ($f_0$) through a driving voltage applied across the gold electrodes. Any material adsorbing or desorbing onto the crystal surface induces a decrease or an increase of the resonance frequency $\Delta f_n = f_n - f_0$ of the n$^{th}$-overtone. $\Delta f_n$ is related to the adsorbed mass per unit area (ng cm$^{-2}$) through the Sauerbrey equation:

$$\Delta m = -C \, \Delta f_n / n \qquad (1)$$

where $C$ is the Sauerbrey constant (17.7 ng s cm$^{-2}$ for a 5 MHz quartz sensor) and $n$ = 1, 3, 5, 7, 9, 11 and 13 is the overtone number. An indication of frictional losses due to viscoelastic properties of the adsorbed layer is provided by changes in dissipation $\Delta D_n = E_D(n)/2\pi E_S(n)$, where $E_D(n)$ is the energy stored in the sensor crystal and $E_S(n)$ is the energy dissipated by the viscous nature of the surrounding medium for the n$^{th}$ overtone.[54,55] The combination of dissipation measurement with the frequency monitoring allows the determination the adsorbed mass as well as layer viscoelastic properties.[56] The Sauerbrey equation mentioned previously assumed that the adsorbed film is laterally homogeneous, evenly distributed and thin, and that the change in resonance frequency is solely due to the adsorbed mass, including water hydrodynamically trapped in the film. However, polymer films are soft, *i.e.* they may exhibit viscoelasticity, and therefore the Voigt model is more appropriate for data treatment. The Voigt model uses frequency and dissipation data from multiple overtones to calculate the thickness, the shear elastic modulus and the shear viscosity of the adsorbed film.[57-59] The model assumes in addition that *i)* the bulk solution above the layer is purely viscous and Newtonian, *ii)* the film is uniform, *iii)* the viscoelastic properties of the layer are frequency independent in the range 5 – 65 MHz, and *iv)* there is no slip between the adsorbed layer and the crystal during shearing.[57] In this work, results at overtones n° 3, 5, 7, 9, 11 and 13 were adjusted to the Voigt model using the QTools software (Biolin Scientific AB, Sweden). The measurements were performed after mounting the crystals in the flow module and establishing a baseline with water. The water was exchanged with PEGylated polymer solution pumped into the chamber. The adsorption behavior of different PEG copolymers at various concentrations on the $Fe_3O_4$ was checked one after another, and DI-water water at pH 7.4 was used to rinse the layer surface between each deposition step. The adsorption behavior of fetal bovine serum (FBS) 10 vol. % on the PEG brush was also studied using the same procedures.

## III - Results and Discussion

### III.1 – Stealth phosphonic acid PEG coated nanoparticles

In this work, 6.8 nm iron oxide nanocrystals were coated with phosphonic and with carboxylic acid PEG copolymers using a formulation pathway described in the Materials and Methods section.[50-52] In brief, dispersions of particles and of PEGylated copolymers were prepared in the same conditions of pH (pH 2.0) and concentration ($c$ = 0.2 wt. %), and mixed at different volume ratios $X$. The pH of the mixed solution was raised to pH 8.0 by sodium hydroxide addition. It was found that below the critical mixing ratio $X_C$, here equal to 1.5 for both polymers well-dispersed coated particles were obtained, with a $D_H$ about 5 nm larger than that of bare particles (Fig. 2a). Above $X_C$ in contrast, particles form large aggregates and precipitate in solution (as uncoated particles do), indicating an incomplete surface coverage from the polymers. The dispersions studied here were prepared at $X$ = 0.2, i.e. with a large excess of polymers to ensure that positive surface charges were all complexed by acid groups. The dispersions were then dialyzed against deionized water to remove the excess polymers (cut-off membrane 50 kD and 100 kD). Dynamic light scattering was used to measure the thickness of the polymer layer. For dispersions that are not stable, light scattering also allows estimating the aggregation kinetics.[46,53] Figs. 2b and 2c display the second autocorrelation function of the scattered intensity $g^{(2)}$(t) for iron oxide coated with phosphonic and carboxylic acid PEG$_{2K}$ copolymers in de-ionized water, respectively. The data exhibit a quasi-exponential decay associated with a unique relaxation mode. Derived from the second cumulant coefficient ($Z_{ave}$), the hydrodynamic diameters were 23.8 and 21.8 nm respectively, with dispersity indexes of 0.08 and 0.18. These $D_H$-values are 9.8 and 7.8 nm larger than that of the bare particles ($D_H$ = 14 nm[24]), and were ascribed to the layer thickness noted $h_{NP}$ to distinguish it from the polymer layer thickness on flat substrate $h_{2D}$ defined in the next section. We found here $h_{NP}$ = 4.5 ± 0.5 nm. With zeta potentials of -2 to -6 mV, electrokinetic measurements confirmed that the PEGylated particles are globally neutral. For PEG$_{5K}$, the polymer thickness was also determined and found at 8 ± 1 nm. The values for $h_{NP}$ are consistent with stretched PEG chains forming a polymer brush.[38,60] As shown in the insets, the associated intensity distributions are characterized by a single population of particle size. When dispersed in 10 vol. % fetal bovine serum, the autocorrelation function and intensity distribution remains unchanged for particles with phosphonic acids ($D_H$ = 21.8 nm, $pdi$ = 0.21). This result ascertains that PEGylated particles are stable in a serum rich medium and devoid of protein corona. A comprehensive characterization study has also shown that this stability is being maintained in cell culture media without serum and for extended period of time (> weeks).[21,24] For particles coated with carboxylic acid functionalized polymers, the $g^{(2)}$(t) relaxation is shifted to longer decay times, and the intensity distribution is now peaked at $D_H$ = 85.3 nm ($pdi$ = 0.21), indicating a modification of the particle structures. The size increase could be due to protein adsorption on the PEG layer, a scenario that would be compliant with the corona model,[3,31,53] or to particle aggregation induced by the PEG layer depletion. From light scattering measurements, it is concluded that phosphonic acid PEG copolymer is an efficient coating agent compared to its carboxylic acid counterpart. Although the impact of the polymer type on particle stability is clear, the nature of the interactions of plasma proteins with a PEG coating layer remains open to question. To answer this question, a series of QCM-D experiments were performed using iron oxide substrates grafted with PEG copolymers.

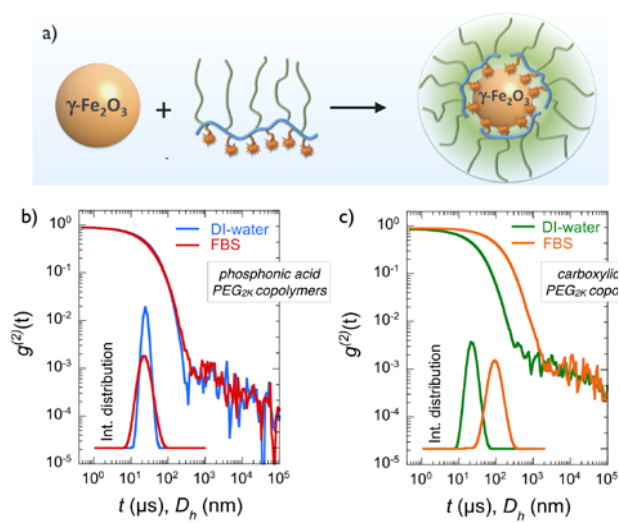

*Figure 2*: a) Schematic representation of an iron oxide particle, of a multi-phosphonic or carboxylic acid PEG copolymer and of the resulting nanostructure made from the two species. b) Autocorrelation function $g^{(2)}$(t) of the scattered light obtained from iron oxide nanoparticles coated with phosphonic acid PEG$_{2K}$ copolymers in DI-water and in fetal bovine serum (FBS) containing medium. Inset: intensity distribution corresponding to the correlograms. c) Same as in Fig. 2b for carboxylic acid PEG$_{2K}$ copolymer coating. The data show the destabilization of the dispersion and the particle agglomeration.

### III.2 – Polymer adsorption on Fe$_3$O$_4$ substrate
*Effect of PEG molecular weight*

Fig. 3a and 3b display the adsorption profiles obtained by means of QCM-D technique for the third overtone frequency $\Delta f_3/3$ and the related dissipation $\Delta D_3$ of a 0.1 wt. % solution of phosphonic acid PEG$_{1K}$ copolymers at pH 2.0 and room temperature, as for nanoparticle coating.[24] Indeed, at this pH Fe$_3$O$_4$ is positively charged, with an estimated density of active Fe-OH$_2^+$ sites of about 1.50 nm$^{-2}$.[47,49] Upon solution injection (arrow at t = 0), the frequency exhibits a rapid decrease and then a fast saturation at $\Delta f_3/3$ = -22.7 Hz. Similarly, $\Delta D_3$ increases rapidly and reaches a plateau at $1.1 \times 10^6$. By increasing PEG molecular weights as in Fig. 3c and 3d for PEG$_{2K}$ and in Fig. 3e and 3f for PEG$_{5K}$, the time dependent profiles remain basically the same but the $\Delta f_3/3$, in absolute values, increases to -29.4 Hz (with a dissipation $\Delta D_3$ = $2.0 \times 10^{-6}$) and -43.3 Hz for PEG$_{2K}$ and -43.3 Hz (with $\Delta D_3$ = $4.8 \times 10^{-6}$) for PEG$_{5K}$. These adsorption kinetics are consistent with those reported by QCM-D on the deposition of polymers on various substrates.[8,41,42,61-64] The polymer binding curves corresponding to the different overtones ($n$ = 3, 5, 7, 9 and 11) are provided in Supplementary Information S7. Fig. 3g summarizes the frequency shift data, *i.e.* the adsorbed mass (including the solvation water) for the three molecular weights, indicating that increasing masses are bound to the substrate depending on chain length.[65] After the deposition, rinsing with DI-water at pH 7.4 has a little effect on the adsorbed layer for the considered polymers. The relative frequency losses after rinsing are 5%, 2% and 1.4% for PEG$_{1K}$, PEG$_{2K}$ and PEG$_{5K}$, respectively, indicating that the polymer layers are firmly attached to the substrate, and that stability is enhanced for longer chains.

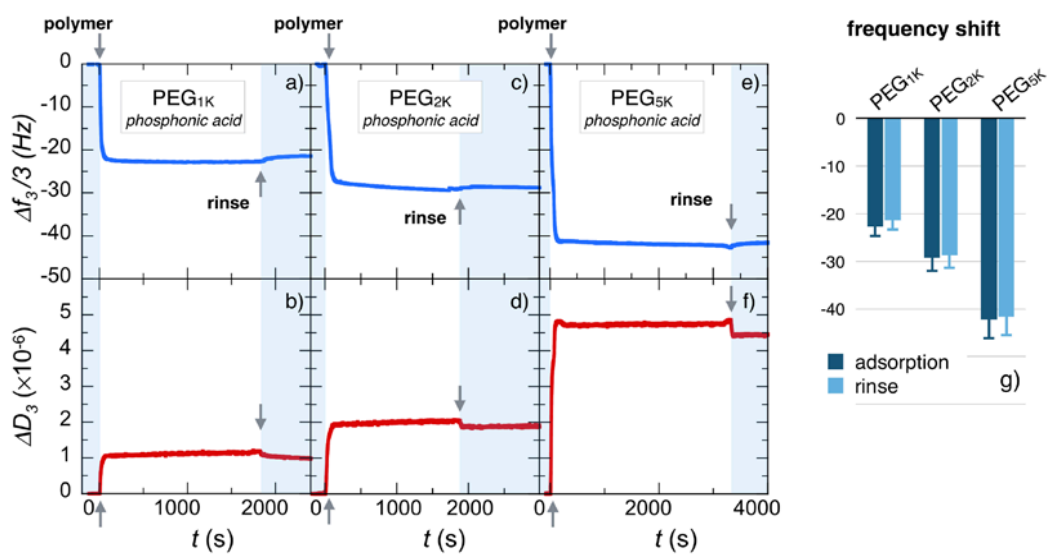

**Figure 3**: Real-time binding curves for frequency $\Delta f_3/3$ and dissipation $\Delta D_3$ during the adsorption of phosphonic acid poly(ethylene glycol) copolymers on Fe$_3$O$_4$ substrates at *pH 2.0*. PEG pending side-chains have molecular weight of 1000 g mol$^{-1}$ (a,b), 2000 g mol$^{-1}$ (c,d) and 5000 g mol$^{-1}$ (e,f). The data are those of the 3$^{rd}$ overtone of the QCM-D acoustic device. In each panel, the first arrow at t = 0 denotes the time at which the polymer solution (concentration 0.1 wt. %) is injected. The second arrow denotes the time at which DI-water (*pH 7.4*) is introduced for rinsing. g) Histogram for the steady state frequencies upon polymer adsorption and rinsing.

*Effect of acid groups*
The study of layer formation of PEG layers, respectively with phosphonic acid and carboxylic acid as linkers, at pH 2.0 sheds light on the role of acidic groups of different strength in the adsorption process. At this acidic pH, indeed phosphonic acid groups are negatively charged, carboxylic acid group are uncharged and the Fe$_3$O$_4$ substrate is positively charged.[47-49] Fig. 4a compares the frequency and dissipation binding curves obtained for PEG$_{2K}$ copolymers functionalized with phosphonic acid and carboxylic acid moieties at pH 2.0 and 25 °C. It can be seen that the carboxylic acid PEG copolymer undergoes a strikingly different adsorption process compared to its phosphonic acid counterpart. In particular, a well defined undershoot behavior is observed and suggests fast adsorption/desorption processes due to conformational rearrangement of the adsorbent on oversaturated surfaces.[66,67] Furthermore, the adsorbed mass at saturation is 30% lower than for phosphonic acid ($\Delta f_3/3$ = - 20.6 Hz *versus* -29.4 Hz), while the related dissipation is slightly higher, globally indicating that the less carboxylic acid PEG copolymers are more loosely bound to the $Fe_3O_4$ substrate than are the phosphonic acid functionalized chains. Accordingly, upon rinsing, the carboxylic acid containing polymer shows a slight decrease and an apparent compaction of the bound mass, which are not observed for with phosphonic acid.

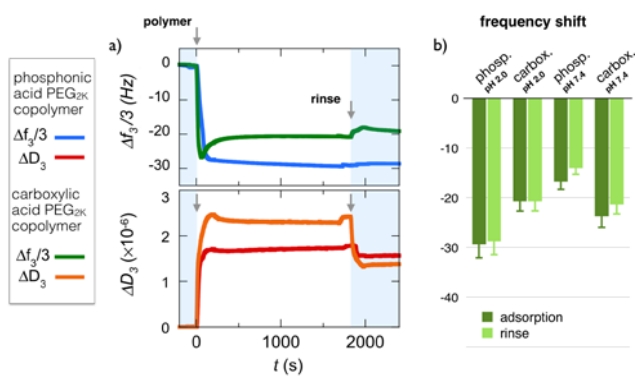

**Figure 4**: a) Binding curves for frequency $\Delta f_3/3$ and dissipation $\Delta D_3$ during the adsorption of phosphonic acid and carboxylic poly(ethylene glycol) copolymers on $Fe_3O_4$ substrates at pH 2.0. Poly(ethylene glycol) side-chains are 2000 g mol$^{-1}$ for both copolymers. In each panel, the first arrow at t = 0 denotes the time at which the polymer solution (concentration 0.1 wt. %) is injected. The second arrow denotes the time at which DI-water (pH 7.4) is introduced for rinsing. b) Histogram for the steady state frequencies upon polymer adsorption and rinsing at the two pH values, pH 2.0 and pH 7.4.

To further assess the role of the charges on adsorption, experiments were performed at pH 7.4, *i.e.* in conditions where the phosphonic and carboxylic acids are both negatively charged and where the $Fe_3O_4$ substrate is neutral.[48,49] Degrees of ionization estimated from $pKa$'s values are 1.3 (indicating that 30% of the phosphonic acid groups bear two negative charges) and 0.8, respectively. Fig. 4b summarizes the frequency shifts for the two acid functionalized PEGs at the two pHs (see Supporting Information S8 for complete adsorption profiles). In particular, the phosphonic acid PEG copolymer suffers a drastic reduction (of about 50%) of the adsorbed mass from pH 2.0 to pH 7.4, while the carboxylic acid functionalized PEG mass is adsorbed in comparable amount at both pHs. Again, rinsing has a negligible effect on the adsorbed masses. Data from Fig. 4b suggest that for phosphonic acid containing polymers, electrostatics is an important driving force for binding iron oxide surface, as adsorption is related to oppositely charge pairing and complexation.[41,63] The lower adsorption levels exhibited by carboxylic acid functionalized PEG may result from the fact that the acid groups and the substrate are only weakly charged. In this later case, other binding mechanism, including H-bonding might be relevant.[10,68,69] In overall, the data indicate that the most efficient coating and binding to the $Fe_3O_4$ substrate occurred in acidic conditions with phosphonic acid groups interacting with protonated Fe-OH$_2^+$ groups. The two different acidic groups then interact with a strikingly different efficiency, depending on the relative pKa's of phosphonic and carboxylic acid residues.

### III.3 – Polymer brush structure

Fig. 5a shows the relationship between the variation of the dissipation $\Delta D_3$ and the resonance frequency shift, $-\Delta f_3/3$, during the adsorption process of phosphonic acid PEG$_{1K}$, PEG$_{2K}$ and PEG$_{5K}$ and carboxylic acid PEG$_{2K}$ at pH 2.0. Previous studies have shown that the slope of the $D - f$ plot reflects the layer viscoelasticity, depending upon processes such as conformational changes, compaction or hydration/dehydration of the macromolecules at the surface.[55,62] For phosphonic acid containing polymers, the ratio $-\Delta D_3/(\Delta f_3/3)$ is found in the range $(0.6 - 1.7) \times 10^{-7}$ Hz$^{-1}$ and remains below the Sauerbrey limit $4 \times 10^{-7}$ Hz$^{-1}$ typical of homogeneous and rigid films.[54,55] For the carboxylic acid PEG$_{2K}$ copolymer, the undershoot observed in Fig. 4a translates into a change of regime with a negative slope 5 minutes after injection. The Sauerbrey equation (Eq. 1) is thus used to derive the areal mass density of the film, which in the present conditions includes both polymer adsorbate and solvent. With increasing molecular weight, from PEG$_{1K}$ to PEG$_{5K}$, the areal mass density increases from 400 to 750 ng cm$^{-2}$ for the deposition step at pH 2.0 (Table 2 and Supporting Information S9), and from 380 to 740 ng cm$^{-2}$ for the rinsing step at pH7.4. Values in the range 200 to 1000 ng cm$^{-2}$ are usual for polymers adsorbing spontaneously at interfaces either *via* physisorption or *via* grafting-to mechanisms.[41,42,61-63]

The hydrodynamic film thickness $h_{2D}$ was estimated using the Voigt viscoelastic model.[54,55] In the deposition step, the thickness increases from 4.1, 5.7 to 9.4 nm for phosphonic acid containing PEG$_{1K}$,

PEG$_{2K}$ and PEG$_{5K}$ polymers respectively (Fig. 5b). The thickness of carboxylic acid functionalized PEG$_{2K}$ is 4.6 nm. After rinsing with DI-water at pH 7.4, $h_{2D}$-values are lowered by 3% – 10%, the reduction being strong for PEG$_{1K}$. Values for PEG$_{2K}$ and PEG$_{5K}$ brushes are in good agreement with those obtained by Nalam *et al.* using poly(L-lysine)-*graft*-PEG[63] and by Emilsson *et al.* using thiol-terminated PEGs.[42] Note also that the thickness of the PEG$_{2K}$ and PEG$_{5K}$ films on flat Fe$_3$O$_4$ substrates compares well with that of the nanoparticle coating layer (Section III.1). From light scattering, spherical brush thicknesses were found at $h_{NP}$ = 4.5 ± 0.5 nm and 8.0 ± 1.5 nm for PEG$_{2K}$ and PEG$_{5K}$, in excellent agreement with the $h_{2D}$ = 5.7 nm and $h_{2D}$ = 9.4 nm measured in QCM-D. The slight difference between the two determinations may arise from curvature effects.[60]

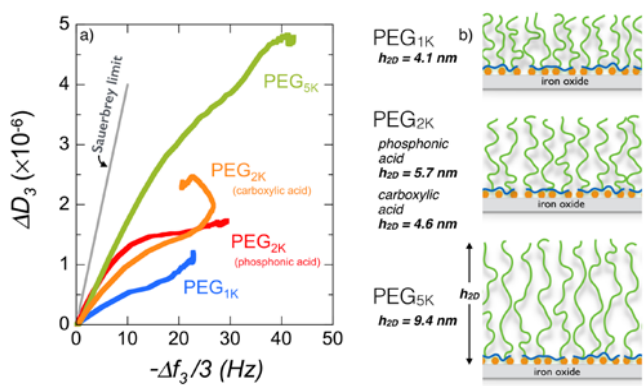

**Figure 5**: a) Plot of the dissipation $\Delta D_3$ as a function of the frequency shift, $-\Delta f_3/3$, during the adsorption process of phosphonic acid PEG$_{1K}$, PEG$_{2K}$ and PEG$_{5K}$ and carboxylic acid PEG$_{2K}$ at pH 2.0. The straight line represents the Sauerbrey limit (slope $4\times10^{-7}$ Hz$^{-1}$) valid for homogeneous and rigid films.[54,55] b) Schematic representation of PEGylated layers deposited on iron oxide substrates obtained for the 4 polymers in a). Also indicated are the values of the brush thicknesses estimated from the Voigt viscoelastic model.[54]

To estimate the PEG density at the surface, it is assumed that the layer structure obeys the polymer brush theory[38,60] in the moderate and high surface density regimes, and that the density $\sigma_{PEG}$ and the height $h_{2D}$ are linked through the relation:[39,40]

$$h_{2D} = \left(\frac{\sigma_{PEG}}{3}\right)^{1/3} b^{2/3} aN \qquad (2)$$

where $a$ = 0.28 nm and $b$ = 0.72 nm are respectively the chemical monomer and Kuhn lengths for poly(ethylene glycol)[29,42] and $N$ the degree of polymerization of the chains. The derived PEG densities are 1.55, 0.57 (0.28) and 0.15 nm$^{-2}$ for PEG$_{1K}$, PEG$_{2K}$ and PEG$_{5K}$, the value in parenthesis being that of carboxylic acid functionalized copolymers. A commonly used parameter for quantitative characterization of polymer brushes is the reduced tethered density $\Sigma = \pi\sigma_{PEG}R_g^2$, where $R_g$ is the gyration radius of the chain in the bulk phase.[34] For poly(ethylene glycol), we adopt recent small-angle neutron scattering results from Le Coeur and coworkers,[70] who found a dependence of the form: $R_g(M_w) = 7.32\times10^{-2}M_w^{0.442}$. From the QCM-D data collected from the different polymers, $\Sigma$ is found in the range 3.8 to 11.7 (Table 2), corresponding to grafting densities in the moderate (carboxylic acid-PEG$_{2K}$, phosphonic acid-PEG$_{5K}$, $1 < \Sigma < 5$) and in the highly stretched regimes (phosphonic acid-PEG$_{1K}$ and PEG$_{2K}$, $\Sigma > 5$).[39] This criterion to evaluate polymer stretching is similar to that found in parallel studies focusing on the ratio $L/2R_g$, where $L$ is the distance between tethered points ($L = \sigma_{PEG}^{-1/2}$).[30,71] A decrease in $\sigma_{PEG}$ by a factor 10 between PEG$_{1K}$ and PEG$_{5K}$ (Tab. II) is attributed to excluded volume interaction and steric repulsion between chains during deposition. The already adsorbed chains act as a barrier to the incoming ones, a mechanism that is more effective for longer chains. As a result, the brush stretching and morphology are different: dense and solid-like for PEG$_{1K}$ and soft and viscoelastic for PEG$_{5K}$.[42] These differences in structure appear also in the different spreading of the overtones measured during deposition (S7, S9).[55] The Voigt viscoelastic model also allows estimating the adsorbed mass from the layer thickness $h_{2D}$.[72] Table S1 in Supporting Information compares the areal mass densities for phosphonic acid PEG copolymers obtained from the Sauerbrey equation (Eq. 1) and from the Voigt model, leading to a good agreement between the two determinations.

In conclusion, phosphonic acid PEG copolymers are shown to adsorb spontaneously at Fe$_3$O$_4$ interfaces and acidic pH. The copolymer backbone attaches to the surface *via* multisite binding, and the PEG side-chains organize themselves into moderate-density or highly stretched brushes of a few nanometers. The QCM-D results confirm the data obtained in the bulk phase with iron oxide nanoparticles.[21]

| terminus | $M_n$ (PEG) g mol$^{-1}$ | mass ng cm$^{-2}$ | thickness $h$ nm | PEG density $\sigma_{PEG}$ nm$^{-2}$ | PEG reduced density $\Sigma$ |
|---|---|---|---|---|---|
| phosphonic acid | 1000 | 400 | 4.1 | 1.55 | 11.7 |
|  | 2000 | 520 | 5.7 | 0.53 | 7.4 |
|  | 5000 | 750 | 9.4 | 0.15 | 4.6 |
| carboxylic acid | 2000 | 370 | 4.6 | 0.28 | 3.8 |

**Table 2:** *Summary of poly(ethylene glycol) copolymers used in QCM-D testing at pH 2.0 and of the parameters describing their brush properties. The mass (ng cm$^{-2}$) is calculated from the Sauerbrey equation (Eq. 1), the layer thickness via the Voigt viscoelastic model, and the PEG density $\sigma_{PEG}$ from the polymer brush theory.[39] The reduced tethered density $\Sigma$ is obtained from the expression $\pi\sigma_{PEG}R_g^2$, where $R_g = 7.32\times10^{-2}M_w^{0.442}$ denotes the PEG gyration radius.[34,70]*

### III.4 – Protein resistance on phosphonated and carboxylated PEG-coated Fe$_3$O$_4$ substrates

We turn now to the protein resistance properties of the PEGylated surfaces. To set a reference, we investigated the QCM-D response of uncoated Fe$_3$O$_4$ substrates exposed to 10 vol. % FBS. As already mentioned, FBS is part of cell culture medium and contains mostly albumin and globulin proteins. Figs. 6a and 6b display the real-time kinetics of protein adsorption in terms of frequency shift $\Delta f_n/n$ and dissipation $\Delta D_n$ for the different overtones (n = 3 – 11). After FBS injection, the frequency of the third mode decreases rapidly and levels off at $\Delta f_3/3 = -57.6$ Hz, whereas in the same time the dissipation increases and reaches $\Delta D_3 = 4.2 \times 10^{-6}$ at steady state. The ratio $-\Delta D_3/(\Delta f_3/3)$ for proteins being lower than the Sauerbrey limit, the frequency data can be translated into the areal mass density, here estimated at 1020 ng cm$^{-2}$ at steady state. As for the polymer adsorption, the QCM-D data show no change in mass density upon rinsing with DI-water at pH 7.4, indicating that proteins are also strongly bound to the Fe$_3$O$_4$ surface (Fig. 6c). The slight decrease in the dissipation is indicative of a deswelling of the protein layer induced by dilution. These findings, together with the values of the protein mass density confirm the strong affinity and resilience of proteins for untreated metal oxide surfaces.[42,56,62]

With PEGylated iron oxide, the adsorption protein behavior changes drastically. In Figs. 7a-7c, $\Delta f_n/n$ and $\Delta D_n$ are plotted for the different overtones following the FBS 10 vol. % injection. Stationary frequency shifts for the third mode are $\Delta f_3/3 = -34.2$ Hz, -14.3 and -11.5 Hz for PEG$_{1K}$, PEG$_{2K}$ and PEG$_{5K}$ respectively, illustrating that the proteins do adsorb on polymer brushes in various amounts. These amounts are however lower than those obtained with untreated surfaces.

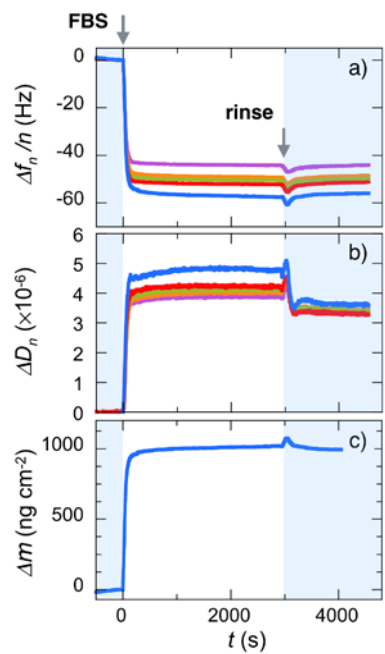

**Figure 6**: *QCM-D responses of uncoated Fe$_3$O$_4$ substrates exposed to a 10 vol. % fetal bovine serum (FBS) solution: a) harmonic resonance frequency $\Delta f_n/n$ with n = 3, 5, 7, 9 and 11, b) corresponding dissipation $\Delta D_n$ and c) areal mass density determined from Eq. 1. The first arrow denotes the FBS injection time and the second arrow the rinsing time.*

After rinsing, the effect is further amplified. For PEG$_{1K}$ layer, the residual frequency shift $\Delta f_{Res}$ is decreased by a factor 3, whereas for PEG$_{2K}$ and PEG$_{5K}$ it is reduced to very low frequency shifts, -1.05 and -0.8 Hz respectively. Related mass densities are 18.6 and 13.6 ng cm$^{-2}$ (Fig. 8a), *i.e.* close to the QCM-D detection limit. For these last samples, results suggest that proteins are only loosely attached to the polymer brush and that rinsing at pH 7.4 washes them up. Another crucial result from Fig. 7 is that the addition of FBS did not modify the PEG$_{2K}$ and PEG$_{5K}$ layer structure, as both $\Delta f_3/3$ and $\Delta D_3$ returned to their pre-injection levels. Fig. 8b displays an histogram of the protein areal mass densities deposited on PEGylated substrates obtained in the conditions of Fig. 4b. Here again, phosphonic acid PEG$_{2K}$ polymers show the best results in terms of protein resistance. A schematic representation of the different adsorption steps for phosphonic acid grafted PEG$_{2,5K}$ layers is shown in Fig. 8c.

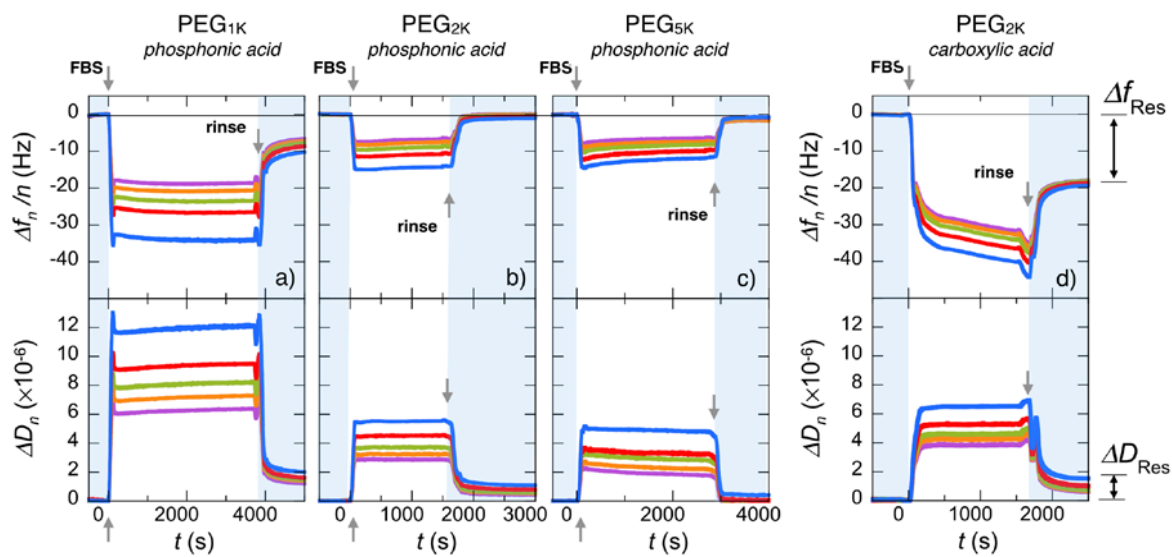

**Figure 7**: Binding curves for frequency $\Delta f_n/n$ and dissipation $\Delta D_n$ (n = 3, 5, 7, 9 and 11) following the injection of a 10 vol. % fetal bovine serum (FBS) solution. a,b,c) Data for phosphonic acid poly(ethylene glycol) coated iron oxide with PEG molecular weight 1000, 2000 and 5000 g mol-1 respectively; d) Same as in Fig. 7b for carboxylic PEG$_{2K}$ copolymers. For each plot, the first arrow denotes the FBS injection time and the second arrow the rinsing time. The residual frequency shift and dissipation after rinsing are noted $\Delta f_{Res}$ and $\Delta D_{Res}$ respectively.

With carboxylic acid PEG copolymers (Fig. 7d), the QCM-D response occurs through a rapid decrease of the frequency shortly after injection, followed by a linear drift of the signal. After 30 min, $\Delta f_3/3$ continues to decrease and reaches -40 Hz that is, three times that found with phosphonic acid functionalized copolymers. Similar behavior is obtained for $\Delta D_3$ that exhibits no saturation plateau. Rinsing shows a rather high residual coverage, associated with a frequency shift of -20 Hz and a dissipation of $\Delta D_3 = 2 \times 10^{-6}$. These findings confirm our hypothesis that carboxylic acid containing polymers are less efficient against protein adsorption. The continuously varying QCM signal following protein injection could indicate a progressive degradation of the PEG layer, due to for instance the competitive complexation between the added proteins and the grafted copolymers. For all polymer studied, the coverage bioresistance can be evaluated from the residual areal mass densities measured on coated and bare surfaces.[8,30,62] The protein adsorption resistance coefficient for phosphonic acid PEG copolymers are 82%, 98% and 99% for PEG$_{1K}$, PEG$_{2K}$ and PEG$_{5K}$, whereas it is only 65% for the carboxylic acid containing polymers. Polymer adsorption at pH 7.4 does not impart resistant coating, as bioresistance percentages are comprised between 60 and 80%. In conclusion, the most repellent layer is that made from phosphonic acid functionalized PEG$_{5K}$ deposited at pH 2.0. As a whole the QCM-D results suggest that the agglomeration of the iron oxide particles coated with carboxylic acid PEG copolymers in FBS 10 vol. % (Fig. 2b) is due to the combined processes of protein adsorption and of brush collapse. This later phenomena are also not present with particles coated with phosphonic acid PEG copolymers.

## IV - Conclusion

In this work, we aim to explain recent results on the pharmacokinetics of sub-10 nm iron oxide particles that are able to travel in the blood compartment of mice for more than 2 hours without being detected by the mononuclear phagocytic system, or cleared by the kidneys or the liver. For *in vivo* assays, the particles were coated with PEGylated polymers specifically synthesized for biomedical applications, and their *in vivo* distribution was monitored by time-resolved magnetic resonance imaging. It should be recalled here that the same core particles without PEG coating have circulation lifetimes of a few minutes, *i.e.* typically 50 times shorter than the PEGylated ones. At the molecular level, it could be shown that this remarkable property is coming from the polymer multi-site attachment at the particle surface and from the adlayer structure. Here we re-examine the issue of nanoparticle stability in protein rich media. For copolymers with identical architecture, and differing by their acidic residues it is demonstrated that the long-term colloidal stability is excellent only with phosphonic acid groups as surface linkers, and that in this case the coated particles are devoid of a protein corona. For the particles grafted with carboxylic acid groups, immediate destabilization of the dispersion is observed, suggesting a protein induced effect towards the coating.

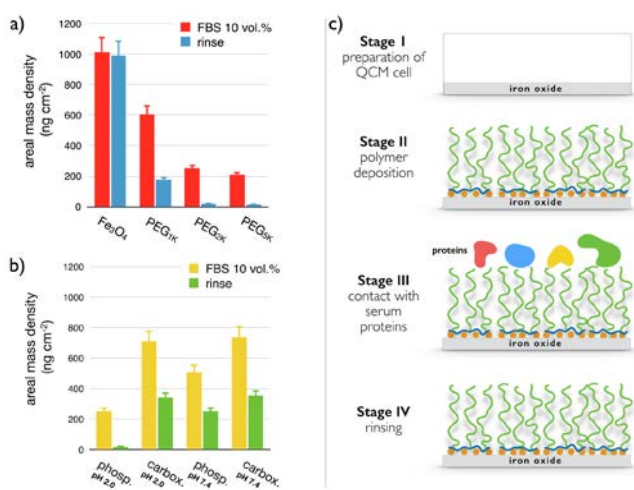

*Figure 8*: a) Protein areal mass densities on PEGylated iron oxide substrates for different PEG molecular weight. b) Same in in a) for different brush formation conditions. With a protein resistance of more than 99%, phosphonic acid PEG$_{5K}$ copolymers deposited at pH 2.0 is the most repellent layer. c) Illustration of the different adsorption and rinsing stages used in QCM-D protocols, including cell preparation (STAGE I), polymer deposition (STAGE II), exposition to serum proteins (STAGE III) and final rinsing (STAGE IV).

To test this hypothesis, QCM-D experiments are performed using iron oxide substrates. In a first investigation, the deposition of different copolymers either with phosphonic acid or carboxylic acid is conducted at different pH conditions and PEG molecular weight. In overall, the areal mass density and the dissipation coefficient indicate that the most efficient binding to Fe$_3$O$_4$ substrates occurs in acidic conditions, with phosphonic residues interacting with protonated Fe-OH$_2^+$ groups. These findings suggest a grafting-to process driven by complexation and by the pairing of electric charges. Estimated from the Voigt model,[55] the hydrodynamic film thickness of flat iron oxide substrate is found to be identical to that of nanoparticles, indicating similar chain conformation in both cases. Comparison with theoretical models allows concluding that the adlayer is forming a polymer brush in the moderate or in the highly stretched regimes. PEG densities ($\sigma_{PEG}$ = 0.15 - 1.5 nm$^{-2}$) and stretching parameters ($\Sigma$ = 3.8 – 11.7) are also estimated.

In a second part, the protein resistance of PEGylated built-up substrates is assessed. The main result that emerges from QCM-D measurement is that whatever the layer extension and grafted PEG density proteins do transiently adsorb on polymer brushes in various amounts. For the PEG$_{2K}$ and PEG$_{5K}$ polymers in combination with phosphonic acid residues however, rinsing is able to wash the proteins away, leading to substrate protein resistance efficacy close to 100%. More importantly, it is found that in the two above cases the addition of fetal bovine serum does not modify the layer structure. In contrast, PEG layers prepared at neutral pH, with PEG$_{1K}$ or from copolymers with carboxylic moieties are globally inefficient against protein adsorption. With such coatings, the protein resistance is mitigated down to 60 – 80%. The present study establishes a clear correlation between the behavior of PEGylated nanoparticles in biological environments, and the protein resistance of PEGylated surfaces made from the same building blocks. In conclusion, in the development of novel formulations for nanomedicine, it is essential that the physicochemical properties of the probes can be predicted to a high degree of accuracy and new approaches must challenge the paradigm of the protein corona. The present work provides answers to these two major questions, confirming that these goals can be achieved thanks to tunable functional polymers.

## Supporting Information

The Supporting Information includes sections on polymer synthesis (S1) and characterization (S2), acid-base titration of the phosphonic acid containing polymers (S3) and NMR characterization (S4). S5 and S6 show TEM data on the iron oxide particles and on the particle stability in biological media respectively.

S7 displays the binding kinetics as a function of the polymer molecular weight for the different overtones whereas S8 compares the phosphonic and carboxylic acid PEG copolymer adsorption as a function of the pH. S9 show the mass kinetic adsorption curve of phosphonic acid poly(ethylene glycol) copolymers on $Fe_3O_4$ substrates. S10 compares the QCM-D data analyzed according the Sauerbrey and Voigt models. This material is available free of charge via the Internet at…

# Acknowledgements

We would like to thank Armelle Baeza, Victor Baldim, Nathalie Mignet, Fanny Mousseau, Evdokia Oikonomou, Chloé Puisney for fruitful discussions. The Laboratoire de PHysico-chimie des Electrolytes et Nanosystèmes InterfaciauX (PHENIX, UMR Université Pierre et Marie Curie-CNRS n° 8234) is acknowledged for providing us with the magnetic nanoparticles. ANR (Agence Nationale de la Recherche) and CGI (Commissariat à l'Investissement d'Avenir) are gratefully acknowledged for their financial support of this work through Labex SEAM (Science and Engineering for Advanced Materials and devices) ANR 11 LABX 086, ANR 11 IDEX 05 02. This research was supported in part by the Agence Nationale de la Recherche under the contract ANR-12-CHEX-0011 (PULMONANO). G.M. and N.G. gratefully acknowledge the financial support to this work from the Project FIRB "Accordi di Programma" (MIUR, Rome, Italy), contract n. RBAP11ZJFA_002.

TOC graphics

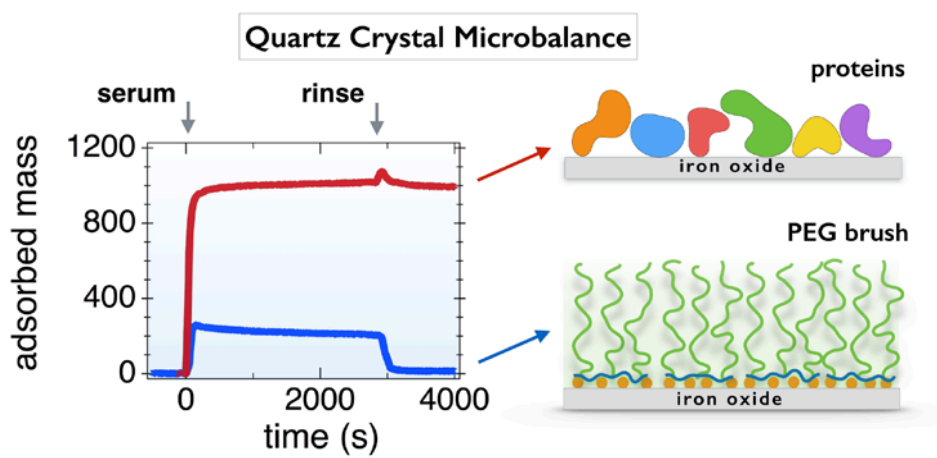

16